\documentclass[aip,reprint,apl,numerical,floatfix]{revtex4-1}
\makeatletter\if@twocolumn\PassOptionsToPackage{switch}{lineno}\else\fi\makeatother

\usepackage{hyperref}
\usepackage{tabulary,graphicx,amsfonts,amsmath,amssymb,amsbsy,booktabs,tabularx}
\usepackage[T1]{fontenc}

\usepackage{url,multirow,morefloats,floatflt,cancel,tfrupee}
\makeatletter

\AtBeginDocument{\@ifpackageloaded{textcomp}{}{\usepackage{textcomp}}}
\makeatother
\usepackage{colortbl}
\usepackage{xcolor}
\usepackage{pifont}
\usepackage[nointegrals]{wasysym}
\makeatletter

\def\mcWidth#1{\csname TY@F#1\endcsname+\tabcolsep}

\def\cAlignHack{\rightskip\@flushglue\leftskip\@flushglue\parindent\z@\parfillskip\z@skip}
\def\rAlignHack{\rightskip\z@skip\leftskip\@flushglue \parindent\z@\parfillskip\z@skip}

\@ifundefined{etal}{}{}

\usepackage{ifxetex}
\ifxetex\else\if@twocolumn\@ifpackageloaded{stfloats}{}{\usepackage{dblfloatfix}}\fi\fi

\AtBeginDocument{
\expandafter\ifx\csname eqalign\endcsname\relax
\def\eqalign#1{\null\vcenter{\def\\{\cr}\openup\jot\m@th
  \ialign{\strut$\displaystyle{##}$\hfil&$\displaystyle{{}##}$\hfil
      \crcr#1\crcr}}\,}
\fi
}

\AtBeginDocument{%
  \@ifpackageloaded{endfloat}%
   {\renewcommand\efloat@iwrite[1]{\immediate\expandafter\protected@write\csname efloat@post#1\endcsname{}}}{\newif\ifefloat@tables}%
}%

\def\BreakURLText#1{\@tfor\brk@tempa:=#1\do{\brk@tempa\hskip0pt}}
\let\lt=<
\let\gt=>
\def\processVert{\ifmmode|\else\textbar\fi}

\@ifundefined{subparagraph}{
\def\subparagraph{\@startsection{paragraph}{5}{2\parindent}{0ex plus 0.1ex minus 0.1ex}%
{0ex}{\normalfont\small\itshape}}%
}{}

\newcommand\role[1]{\unskip}
\newcommand\aucollab[1]{\unskip}
  
\@ifundefined{tsGraphicsScaleX}{\gdef\tsGraphicsScaleX{1}}{}
\@ifundefined{tsGraphicsScaleY}{\gdef\tsGraphicsScaleY{.9}}{}
\def\checkGraphicsWidth{\ifdim\Gin@nat@width>\linewidth
	\tsGraphicsScaleX\linewidth\else\Gin@nat@width\fi}

\def\checkGraphicsHeight{\ifdim\Gin@nat@height>.9\textheight
	\tsGraphicsScaleY\textheight\else\Gin@nat@height\fi}

\def\fixFloatSize#1{}
\let\ts@includegraphics\includegraphics

\def\inlinegraphic[#1]#2{{\edef\@tempa{#1}\edef\baseline@shift{\ifx\@tempa\@empty0\else#1\fi}\edef\tempZ{\the\numexpr(\numexpr(\baseline@shift*\f@size/100))}\protect\raisebox{\tempZ pt}{\ts@includegraphics{#2}}}}

\AtBeginDocument{\def\includegraphics{\@ifnextchar[{\ts@includegraphics}{\ts@includegraphics[width=\checkGraphicsWidth,height=\checkGraphicsHeight,keepaspectratio]}}}

\DeclareMathAlphabet{\mathpzc}{OT1}{pzc}{m}{it}

\def\URL#1#2{\@ifundefined{href}{#2}{\href{#1}{#2}}}

\def\UrlOrds{\do\*\do\-\do\~\do\'\do\"\do\-}%
\g@addto@macro{\UrlBreaks}{\UrlOrds}

\edef\fntEncoding{\f@encoding}

\makeatother

\newif\ifmultipleabstract\multipleabstractfalse%
\usepackage{float,newfloat}

\begin{document}

\def\authorCount{2}
\def\affCount{2}

\def\journalTitle{Applied Physics Letters}

\title{Theoretical Investigation of Performance-Improved Ferroelectric Tunnel Junction Based on Trap-Assisted Tunneling}
\author{Shi-Xi Kong}
\email{281611631kong@gmail.com}
\affiliation{Department of Electronics and Electrical Engineering\unskip, National Yang Ming Chiao Tung University\unskip, 30010\unskip, Hsinchu\unskip, Taiwan, Republic of China}
\author{Tuo-Hung Hou}
\email{thhou@nycu.edu.tw}
\affiliation{Department of Electronics and Electrical Engineering\unskip, National Yang Ming Chiao Tung University\unskip, 30010\unskip, Hsinchu\unskip, Taiwan, Republic of China}
\affiliation{Institute of Electronics\unskip, National Yang Ming Chiao Tung University\unskip, Hsinchu\unskip, 30010\unskip, Taiwan, Republic of China}

\begin{abstract}
CMOS-compatible HfO$_2$-based ferroelectric tunnel junction (FTJ) has attracted significant attention as a promising candidate for in-memory computing (IMC) due to its extremely low power consumption. However, conventional FTJs face inherent challenges that hinder their practical applications. Insufficient current density and limited on-off current ratios in FTJs are primarily constrained by their dependence on direct and Fowler-Nordheim tunneling mechanisms. Building on previous experimental results, this paper proposes a trap-assisted tunneling (TAT)-based FTJ that leverages the TAT mechanism to overcome these limitations. A comprehensive FTJ model integrating ferroelectric switching, direct, Fowler-Nordheim tunneling, and TAT mechanisms is developed, enabling detailed analyses of the trap conditions and their impact on performance. Through systematic optimization of trap parameters and device structure, the simulated TAT-based FTJ achieves ultra-high current density and a remarkable on-off current ratio, meeting the nanoscale IMC requirements. The results highlight the potential of TAT-based FTJs as high-performance memory solutions for IMC applications.
\end{abstract}\def\keywordstitle{Keywords}

\maketitle 

Ferroelectric tunnel junction (FTJ) is a non-volatile memory that allows non-destructive readout between opposite ferroelectric polarization states through the tunnel electroresistance (TER) effect.\cite{ref1,ref2} Due to its extremely low power consumption, CMOS-compatible HfO$_2$-based FTJ has gained significant attention as a promising candidate for in-memory computing (IMC).\cite{ref3,ref4,ref5} To balance the power and the latency, a desirable memory cell for IMC requires on-state resistance in the range of 100~k$\Omega$\,$\sim$\,1~M$\Omega$.\cite{ref6,ref7} As the device area is scaled down to nanoscale, such as a $10\ \mathrm{nm}\times10\ \mathrm{nm}$ FTJ cell and a read voltage of 0.1 V, achieving an ultra-high on-current density ($J_{\mathrm{ON}}$) exceeding $10^5$ A/cm$^2$ becomes imperative. Additionally, a high on-off current ratio ($J_{\mathrm{ON}}/J_{\mathrm{OFF}}$) and low-voltage operation are crucial to enabling high-performance IMC.\cite{ref6,ref7}

To date, conventional FTJs have predominantly relied on direct and Fowler-Nordheim tunneling mechanisms, as shown in Fig.~1(a).\cite{ref8} However, these mechanisms face critical limitations in IMC applications. First, the $J_{\mathrm{ON}}$ falls far below the required $10^5$ A/cm$^2$, even with aggressive thickness scaling down to the ultra-thin range (1--3 nm), achieving $J_{\mathrm{ON}}$ above $10^2$ A/cm$^2$ remains elusive.\cite{ref9,ref10,ref11} Second, while incorporating dielectric materials such as SiO$_2$ or Al$_2$O$_3$ as an interfacial layer (IL) has been explored to enhance the $J_{\mathrm{ON}}/J_{\mathrm{OFF}}$ ratio, exceeding a ratio of 100$\times$ remains a considerable challenge.\cite{ref12,ref13} Moreover, attempts to improve either $J_{\mathrm{ON}}$ or the $J_{\mathrm{ON}}/J_{\mathrm{OFF}}$ ratio often result in inherent trade-offs, limiting the performance of conventional FTJs.\cite{ref2} Interestingly, Chu et al.\ experimentally reported that a 3 nm-thick HZO FTJ with an additional 1.5 nm-thick top IL (TIL) demonstrated simultaneous increases in both $J_{\mathrm{ON}}$ and $J_{\mathrm{ON}}/J_{\mathrm{OFF}}$, as shown in Fig.~1(b).\cite{ref11} This unexpected enhancement in $J_{\mathrm{ON}}$ with a thicker TIL cannot be explained by direct and Fowler-Nordheim tunneling mechanisms alone and was tentatively attributed to traps introduced by the Al$_2$O$_3$ layer.\cite{ref11} However, the microscopic mechanism behind this observation remains unclear and even counter-intuitive, since traps are often associated with leakage currents.\cite{ref14,ref15} Therefore, investigating whether and under what conditions traps can lead to the concurrent enhancement of $J_{\mathrm{ON}}$ and $J_{\mathrm{ON}}/J_{\mathrm{OFF}}$ is essential for overcoming the limitations of conventional FTJs and advancing their applications in IMC.

\bgroup
\fixFloatSize{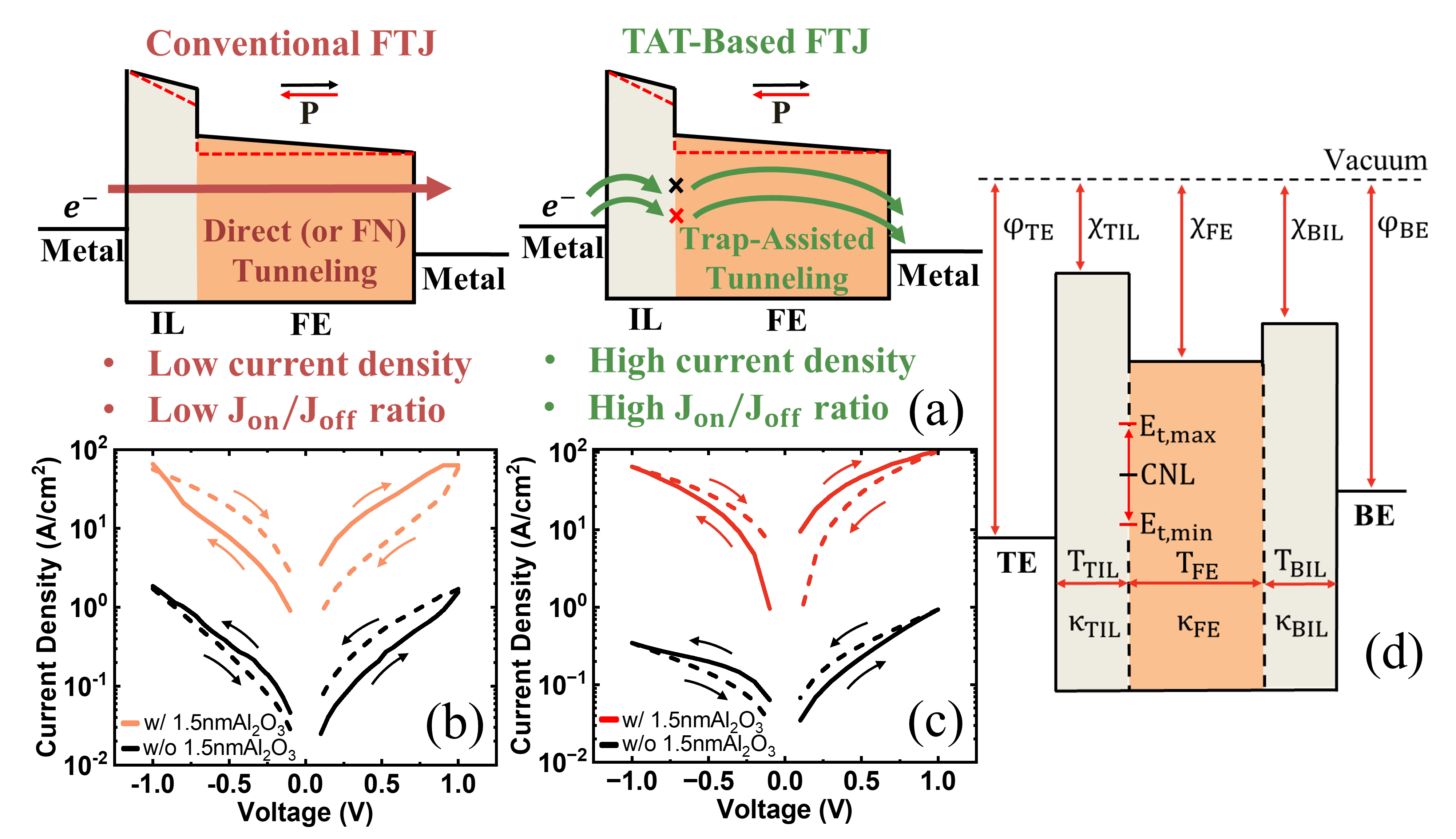}
\begin{figure*}[!htbp]
\centering \makeatletter\IfFileExists{images/Figure1.png}{\includegraphics{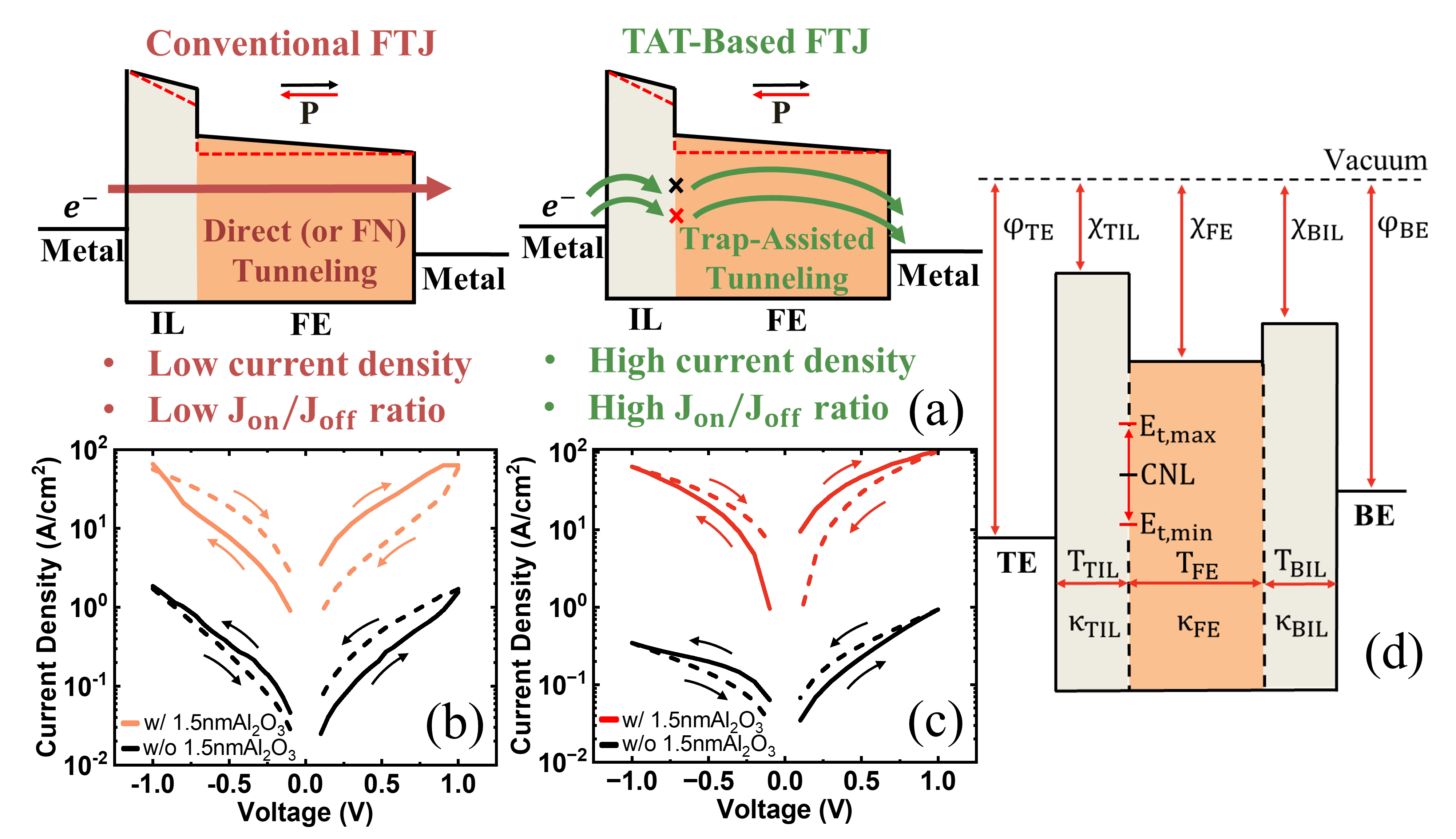}}{\includegraphics{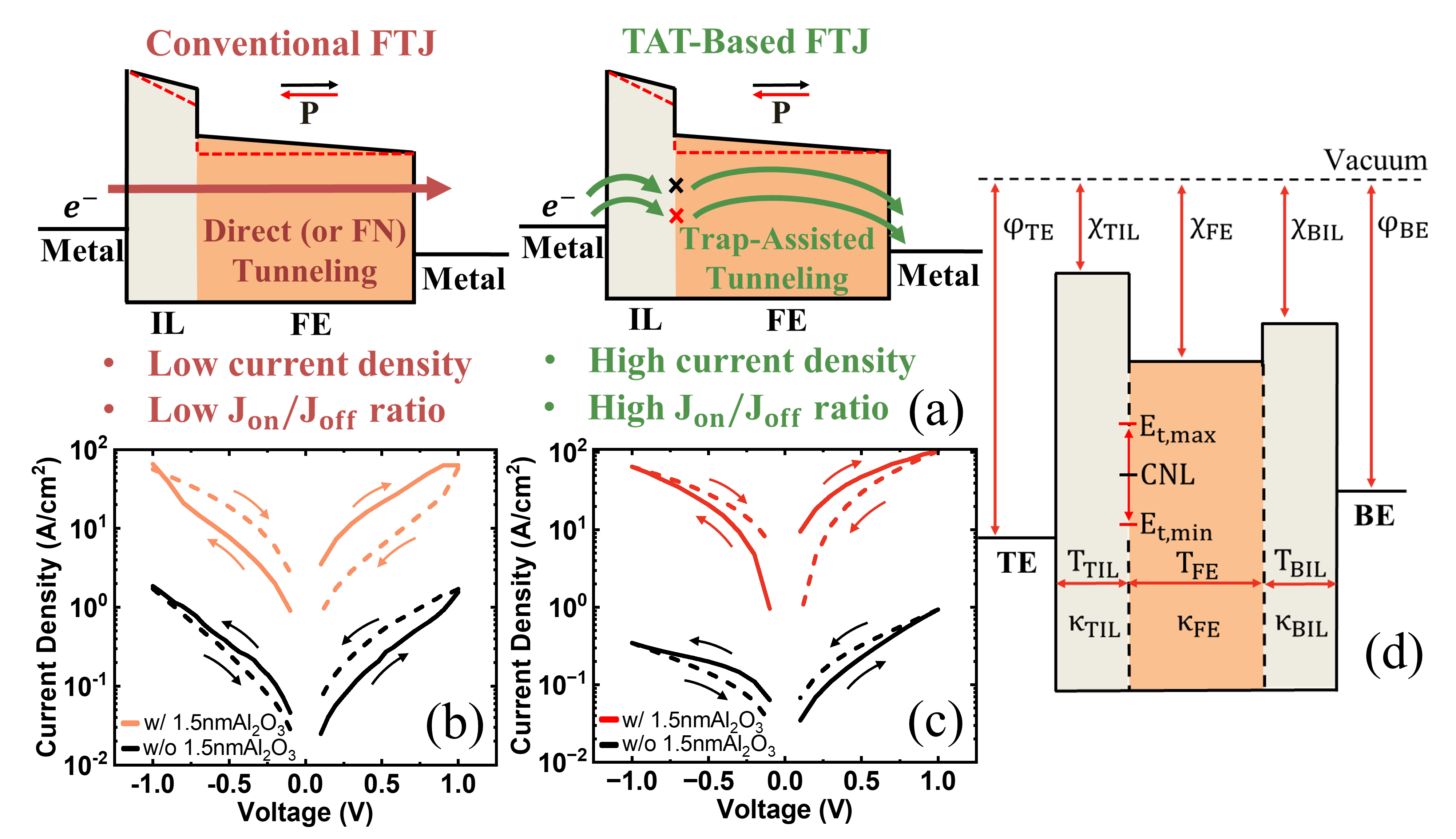}}
\makeatother
\caption{{(a) Comparative schematic of conventional and TAT-based FTJ mechanisms. Conventional FTJ suffers from low \ensuremath{J_{\mathrm{ON}}} and low $J_{\mathrm{ON}}/J_{\mathrm{OFF}}$ ratio, whereas TAT-based FTJ achieves high \ensuremath{J_{\mathrm{ON}}} and high $J_{\mathrm{ON}}/J_{\mathrm{OFF}}$ ratio through appropriate trap modulation. (b) \textbf{Reproduced experimental} and (c) \textbf{simulated} J-V curves w/ and w/o 1.5 nm Al\ensuremath{_{2}}O\ensuremath{_{3}} in 3 nm HZO FTJ, with sweep voltage $\pm1\;\mathrm{V}$. (d) The schematic of the device structure and trap conditions of the simulated FTJ.}}
\label{Figure1}
\end{figure*}
\egroup

In this work, a TAT-based FTJ, where the trap-assisted tunneling (TAT)\cite{ref16,ref17} serves as the dominant tunneling mechanism, is proposed, as shown in Fig.~1(a). Various trap conditions and device structures are investigated through physical modeling, and guidelines for optimizing TAT-based FTJ are provided.

To perform a comprehensive theoretical investigation of TAT-based FTJ, an improved FTJ model was developed, integrating FE switching, direct, Fowler-Nordheim, and TAT mechanisms. Figure~1(d) schematically illustrates the Ni/Al$_2$O$_3$/HZO/BIL/TiN MIFIM-FTJ structure. 
The simulated stack follows the experimental configuration reported by Chu \textit{et al.}~\cite{ref11}, in which a bottom interfacial layer (BIL) is observed at the HZO/TiN interface in transmission electron microscopy images. Previous studies have attributed this BIL to either monoclinic-phase HZO~\cite{m-phaseBIL} or TiN$_x$O$_y$ formation~\cite{TiNxOyBIL}. In this work, it is modeled as monoclinic-phase HZO.

The FE switching, trap dynamics, and electric potential are calculated through an iterative self-consistent scheme. The P--V loop is defined by the Preisach model.\unskip~\cite{S1} The FE thickness-dependent saturation polarization ($P_\mathrm{s}\propto T_{\mathrm{FE}} $), remnant polarization ($P_\mathrm{r}\propto T_{\mathrm{FE}} $), and coercive field ($E_\mathrm{c}\propto T_{\mathrm{FE}}^{-0.61} $) are considered~\cite{ref23,S3}.
Please refer to S1 of the supplementary material for more details on the major and minor P--V loop modeling.

The trap dynamics are evaluated based on the framework proposed by Guan \textit{et al.}~\cite{ref16}, which describes a phonon-assisted inelastic TAT mechanism in HfO$_2$ to capture the lattice-mediated energy dynamics in defective oxides. In a quasi-steady state, the current continuity equation is in the form of
\begin{multline}
(1-f_n)\!\left(R_n^{iT}+R_n^{iB}\right)
 -f_n\!\left(R_n^{oT}+R_n^{oB}\right) \\
 +\sum_{m\neq n}\!\!\left[(1-f_n)R_{mn}f_m
   -f_n R_{nm}(1-f_m)\right]=0 .
\end{multline}
$f_n\in[0,1]$ is the electron occupation probability. $R_{mn}$ is the hopping rate from trap $m$ to $n$, which can be described by the Mott hopping model.~\cite{ambegaokar1971hopping, kopperberg2021consistent} $R^{oT(B)}_n$ denotes the electron hopping rate from trap $n$ to the top (bottom) electrode, and $R^{iT(B)}_n$ represents the hopping rate from the top (bottom) electrode to trap $n$, both are calculated using the Wentzel–Kramers–Brillouin (WKB) approximation~\cite{S7}.
The self-consistent calculation is performed through two iterative loops. 
The outer loop iterates the FE switching, and the inner loop iterates the trap dynamics and potential. 
Initially, in the outer loop, the voltage across the FE layer, $V_{\mathrm{FE}}$, is guessed under a particular applied voltage. 
Subsequently, an initial electron occupation probability $f_n^{\mathrm{ini}}$ is assumed in the inner loop. 
The electric potential $V(x)$ is then calculated by solving the 1D Poisson equation including FE bound charge and trap charge. 
The occupation $f_n$ is then updated through Eq.~(2). 
This iteration continues until the occupation variation $\Delta f_n = |f_n^{\mathrm{out}} - f_n^{\mathrm{ini}}|$ converges below a predefined tolerance $\epsilon_1$. 
The converged potential is then fed back to the outer loop to update $V_{\mathrm{FE}}$, following the same convergence condition $\Delta V_{\mathrm{FE}} \le \epsilon_2$, ensuring a self-consistent solution. 
The applied voltages at the turning points are first self-consistently obtained to construct the minor P--V loop from Eq.~(1), after which the remaining bias points are calculated accordingly.

Based on previous self-consistent calculations, the current density is then calculated. Fowler-Nordheim tunneling is similar to direct tunneling, representing the same quantum-tunneling phenomenon calculated using the Tsu–Esaki model~\cite{tsu1973tunneling} as
\begin{equation}
J_{\mathrm{DT/FN}} = \frac{4\pi m^\ast q}{h^3}
\int T(E)\,\mathcal{N}(E)\, dE,
\end{equation}
where \(T(E)\) is the transmission probability obtained from the self-consistent potential by WKB approximation~\cite{S7}, and \(\mathcal{N}(E)\) is the supply function determined by the Fermi distributions of the top and bottom electrodes.

And the TAT current is obtained from the net electron transfer between traps and top (or bottom) electrodes as
\begin{equation}
J_{\mathrm{TAT}} = -\frac{q}{A} \sum_{n}\Big[(1 - f_n) R_n^{i\,T} - f_n R_n^{o\,T}\Big],
\end{equation}
where A is the cross-sectional area of FTJ.
The total current is therefore determined by the sum of direct/Fowler-Nordheim tunneling and TAT contributions, i.e.,
\begin{equation}
J_{\mathrm{total}} = J_{\mathrm{DT/FN}} + J_{\mathrm{TAT}}.
\end{equation}

In this work, the electron affinities $\chi$ are set to 1.58~eV and 2.2~eV for Al$_2$O$_3$~\cite{Al2O3EA} and HZO~\cite{HZOEA}, respectively. 
The work function $\phi$ of Ni is 5.15~eV~\cite{NiWF}, while TiN exhibits a reported range of 4.1--5.3~eV~\cite{TiNWF}. For simplicity, a representative value of 5.15~eV is adopted. 
The dielectric constants $\kappa$ are 9 for Al$_2$O$_3$~\cite{Al2O3DEconstant} and 30/22 for the ferroelectric orthorhombic and non-ferroelectric monoclinic phases of HZO~\cite{HZOophaseDEconstant,HZOmphaseDEconstant}, corresponding to the FE layer and the BIL, respectively. 
The trap sites are assumed to be located at the Al$_2$O$_3$/HZO interface for simplicity in modeling, where interfacial defects are expected to preferentially form during film deposition at the heterogeneous interface.
The trap energy levels ($E_{\mathrm{T}}$) are assumed to be continuously distributed in 1.1–2.9~eV below the conduction band of the FE layer, following the range reported for HfO$_2$~\cite{HfO2trapenergyrange}, which lies energetically closer to the Fermi level than that of Al$_2$O$_3$ (1.7–2.0~eV below the conduction band of the TIL)~\cite{Al2O3trapenergyrange}. A charge-neutrality level (CNL) is introduced, above which the traps behave as acceptor-like and below which they act as donor-like states. The CNL is set in the middle of the energy range. Its exact value is not critical, as it only shifts the pinning level without altering the overall pinning behavior discussed later.
The major $P$–$V$ loop and the trap surface density $N_T$ were determined by fitting the experimental data. 
Table~I summarizes the extracted fitting parameters and the layer thicknesses of the 1.5-nm Al$_2$O$_3$/3-nm HZO/0.6-nm BIL and the 3-nm HZO/0.6-nm BIL FTJs. As shown in Fig.~1(c), the simulation result, which replicates the experimental conditions, shows excellent agreement with that reported by Chu et al.\cite{ref11} (For more details on the modeling method, see Sec.~S1 of the supplementary material.)

\begingroup
\makeatletter
\long\def\@makecaption#1#2{%
  \vskip\abovecaptionskip
  #1.\ #2\par
  \vskip\belowcaptionskip}%
\makeatother
\begin{table}[!h]
\caption{\raggedright The extracted fitting parameters and the layer thicknesses of the 1.5-nm Al\textsubscript{2}O\textsubscript{3}/3-nm HZO/0.6- nm BIL and the 3-nm HZO/0.6-nm BIL FTJ.}
\centering
\renewcommand{\arraystretch}{1.5}
\setlength{\tabcolsep}{1pt}
\begin{tabular}{@{} >{\raggedright\arraybackslash}m{2.4cm} >{\raggedright\arraybackslash}m{3.2cm} >{\raggedright\arraybackslash}m{3cm} @{}}
\toprule
Symbol          & w/ 1.5nm Al\textsubscript{2}O\textsubscript{3} & w/o 1.5nm Al\textsubscript{2}O\textsubscript{3} \\ 
\midrule
$k_s(= P_s/T_\mathrm{FE})$     & $3\;\mu\mathrm{C}/\mathrm{cm}^2 \cdot \mathrm{nm}$ & $3\;\mu\mathrm{C}/\mathrm{cm}^2 \cdot \mathrm{nm}$     \\
$k_r(= P_r/T_\mathrm{FE})$     & $2.997\;\mu\mathrm{C}/\mathrm{cm}^2 \cdot \mathrm{nm}$ & $2.997\;\mu\mathrm{C}/\mathrm{cm}^2 \cdot \mathrm{nm}$     \\
$k_c(= E_c/T_\mathrm{FE}^{0.61})$ & $2.5\;\mathrm{MV}/\mathrm{cm} \cdot \mathrm{nm}^{-0.61}$ & $2.5\;\mathrm{MV}/\mathrm{cm} \cdot \mathrm{nm}^{-0.61}$ \\
$N_\mathrm{T}$           & $10^{12}\;\mathrm{cm}^{-2}$              & $0\;\mathrm{cm}^{-2}$ \\
$T_\mathrm{TIL}$         & $1.5\;\mathrm{nm}$                          & $0\;\mathrm{nm}$ \\
$T_\mathrm{FE}$         & $3\;\mathrm{nm}$                            & $3\;\mathrm{nm}$  \\
$T_\mathrm{BIL}$         & $0.6\;\mathrm{nm}$                          & $0.6\;\mathrm{nm}$ \\
\bottomrule
\end{tabular}
\label{table:values}
\end{table}
\endgroup

\bgroup
\fixFloatSize{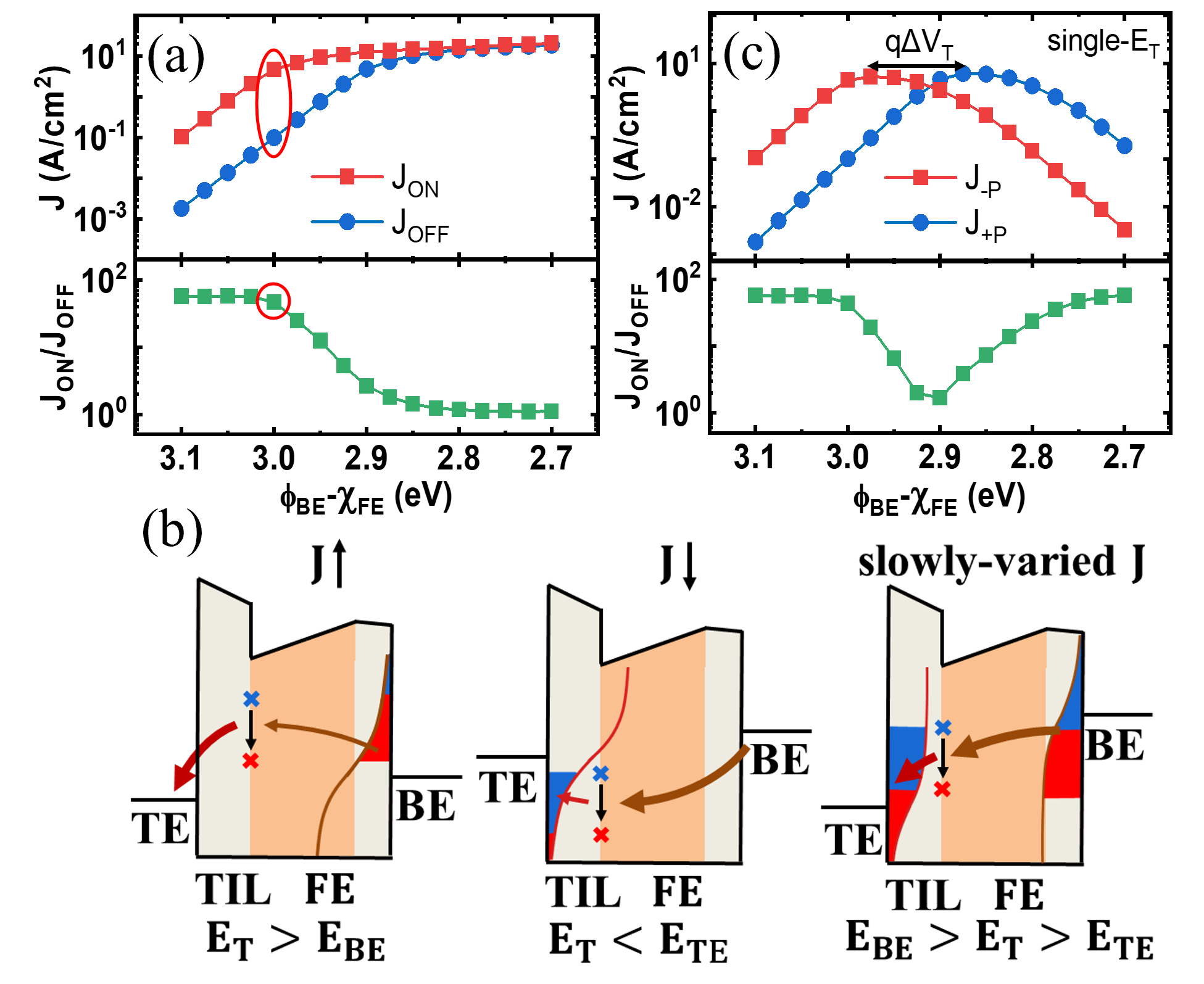}
\begin{figure*}[!htbp]
\centering \makeatletter\IfFileExists{images/Figure2.png}{\includegraphics{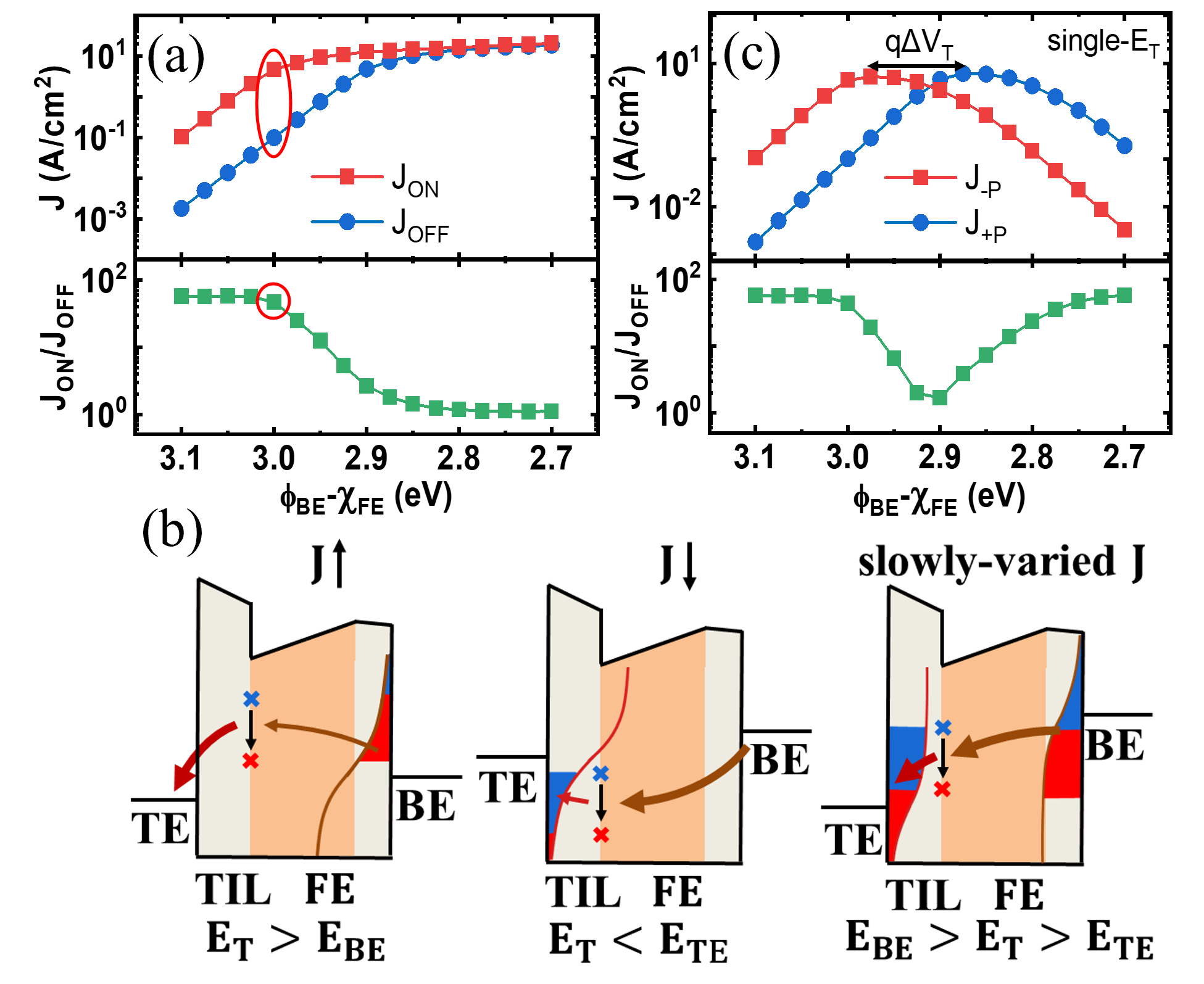}}{\includegraphics{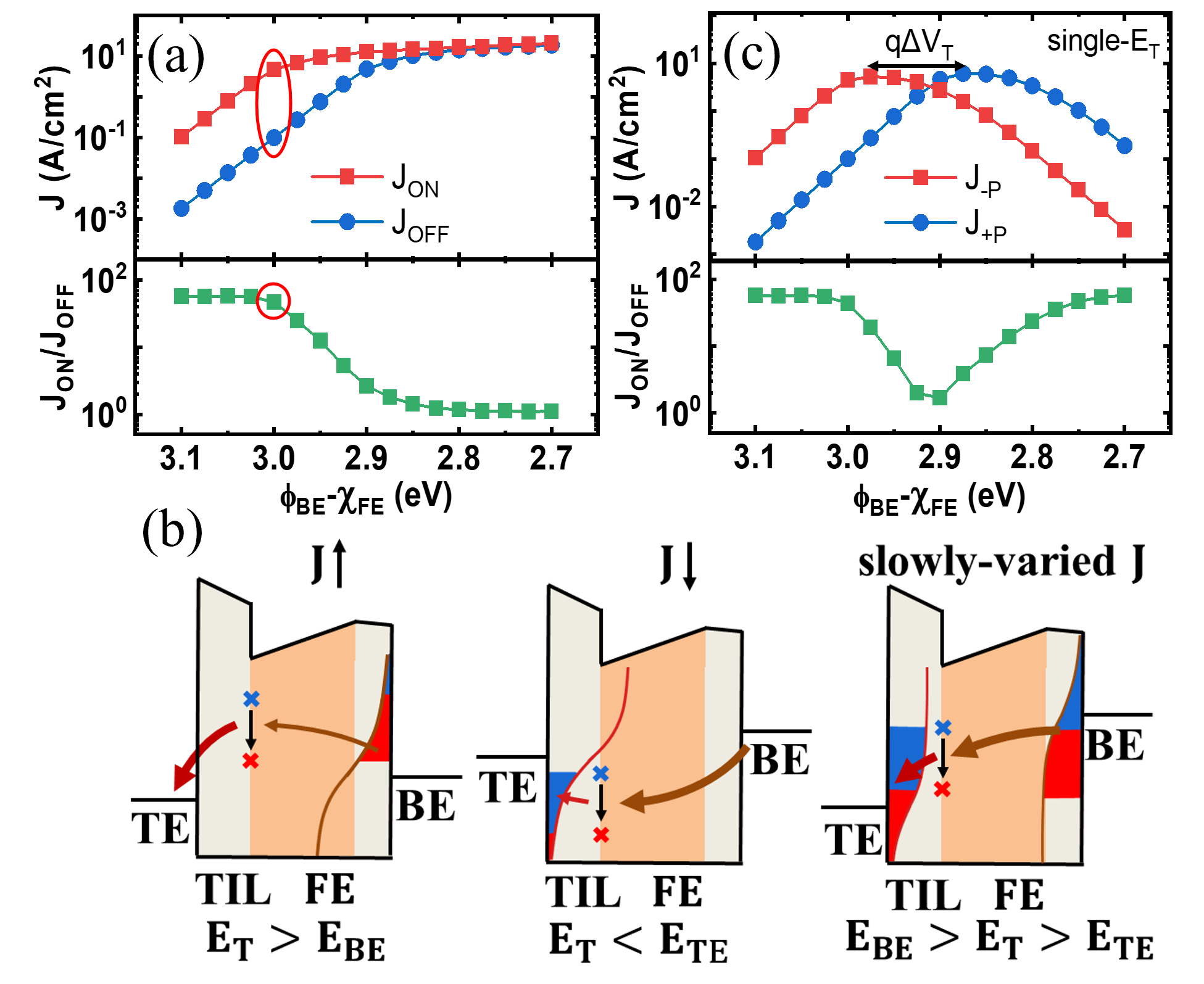}}
\makeatother 
\caption{{(a) \ensuremath{E_{\mathrm{T}}} dependence of \ensuremath{J_{\mathrm{TAT}}} under $V_{\mathrm{write}}=1\;\mathrm{V}$ and $V_{\mathrm{read}}=0.1\;\mathrm{V}$. (b) Schematic illustration of the single \ensuremath{E_{\mathrm{T}}} dependence of \ensuremath{J_{\mathrm{TAT}}} under positive \ensuremath{V_{\mathrm{read}}}, with \ensuremath{E_{\mathrm{TE}}} and \ensuremath{E_{\mathrm{BE}}} being the Fermi level of TE and BE. (c) Single \ensuremath{E_{\mathrm{T}}} dependence of \ensuremath{J_{\mathrm{TAT}}} under $V_{\mathrm{write}}=1\;\mathrm{V}$ and $V_{\mathrm{read}}=0.1\;\mathrm{V}$.}}
\label{Figure2}
\end{figure*}
\egroup

Based on the experimental structure, the impact of trap conditions is investigated in the 1.5-nm Al$_2$O$_3$/3-nm HZO/0.6-nm BIL FTJ. Figure~2 presents the impact of trap energy level. A practical approach to modulating trap behavior involves adjusting the FE band offset ($\varphi_{\mathrm{BE}} - \chi_{\mathrm{FE}}$). As shown in Fig.~2(a), the simulations reveal a balance point between $J_{\mathrm{ON}}$ and $J_{\mathrm{ON}}/J_{\mathrm{OFF}}$, where a higher FE band offset significantly reduces $J_{\mathrm{ON}}$, while a lower offset rapidly degrades $J_{\mathrm{ON}}/J_{\mathrm{OFF}}$.

This performance dependence on trap energy level can be explained as the cumulative effect of multiple single-trap scenarios. Figure~2(b) illustrates the dependence of $J_{\mathrm{TAT}}$ on a single $E_{\mathrm{T}}$ under positive read voltage. For $E_{\mathrm{T}}$ larger than the Fermi level of the bottom electrode (BE), the slower BE-to-trap tunneling dominates the BE-to-TE TAT process, causing $J_{\mathrm{TAT}}$ to increase as $E_{\mathrm{T}}$ decreases, driven by the exponential increase of carrier injection governed by the Fermi distribution. Conversely, for $E_{\mathrm{T}}$ smaller than the Fermi level of the top electrode (TE), trap-to-TE tunneling dominates, and $J_{\mathrm{TAT}}$ decreases with decreasing $E_{\mathrm{T}}$ due to the exponential decrease in carrier filling. In the intermediate range where $E_{\mathrm{T}}$ lies between the Fermi level of TE and BE, both BE-to-trap and trap-to-TE tunneling occur from higher to lower energy levels. In this region, carrier injection and filling no longer change exponentially, resulting in a gradual evolution of $J_{\mathrm{TAT}}$ with $E_{\mathrm{T}}$.

The findings in Fig.~2(b) indicate that, in the TAT-based FTJ, the TER effect is governed by distinct $E_{\mathrm{T}}$ under opposite polarization states. Figure~2(c) illustrates the performance dependence on single $E_{\mathrm{T}}$, which is set at 2.9~eV below the conduction band of FE layer. $J$ under positive and negative FE polarization ($J_{+\mathrm{P}}$ and $J_{-\mathrm{P}}$) exhibit similar trends but are offset by a $E_{\mathrm{T}}$ difference, which is proportional to the potential difference at the TIL–FE interface. Consequently, the $J_{\mathrm{ON}}/J_{\mathrm{OFF}}$ ratio increases exponentially with this potential difference, mirroring the trend between $J_{\mathrm{TAT}}$ and $E_{\mathrm{T}}$ discussed earlier, and achieves a much higher value than that
 caused by direct or Fowler–Nordheim tunneling. However, the $J_{\mathrm{ON}}/J_{\mathrm{OFF}}$ ratio sharply degrades when $E_{\mathrm{T}}$ under positive or negative FE polarization falls within the intermediate range. Therefore, the optimal condition for trap energy level under a positive read voltage is achieved when the lowest energy level under negative FE polarization aligns with the Fermi level of BE, ensuring no traps reside in the intermediate range. Similarly, under a negative read voltage, the optimal condition occurs when the lowest energy level in the under positive FE polarization distribution aligns with the Fermi level of TE.

After elucidating the role of trap energy level, Fig.~3(a) and 3(b) present the impact of trap surface density. As shown in Fig.~3(a), when $N_{\mathrm{T}}$ is small, direct/Fowler-Nordheim tunneling dominates, and the device behaves as a conventional FTJ. As $N_{\mathrm{T}}$ increases, TAT begins to dominate, leading to higher $J_{\mathrm{ON}}$ without degrading the $J_{\mathrm{ON}}/J_{\mathrm{OFF}}$ ratio, as $J_{\mathrm{TAT}}$ is proportional to $N_{\mathrm{T}}$.

However, this result holds only when the trap charge ($Q_{\mathrm{T}}$) has limited effects on the local electric potential. Once $N_{\mathrm{T}}$ exceeds a threshold, the $J_{\mathrm{ON}}/J_{\mathrm{OFF}}$ ratio diminishes rapidly. This arises from the Fermi level pinning of the charge neutrality level, a phenomenon that has been extensively studied in Schottky diodes\cite{ref19,ref20}, and has also been extended to dielectric interfaces such as HfO$_2$/SiO$_2$.\cite{ref21} CNL is defined as the energy level that must be filled to achieve charge neutrality at the interface and distinguishes between donor-like and acceptor-like traps.~\cite{ref22} When CNL is larger than the Fermi level at the traps, partially unfilled donor-like traps result in a net positive $Q_{\mathrm{T}}$. When CNL is smaller than the Fermi level at the traps, partially filled acceptor-like traps result in a net negative $Q_{\mathrm{T}}$. As shown in Fig.~3(b), as $N_{\mathrm{T}}$ increases, CNL in both the on-state ($\mathrm{CNL}_{\mathrm{ON}}$) and off-state ($\mathrm{CNL}_{\mathrm{OFF}}$) eventually merges with the Fermi level at the traps, positioned between the Fermi level of TE and BE. At high $N_{\mathrm{T}}$, even small energy shifts in CNL produce significant changes in $Q_{\mathrm{T}}$, pinning the CNL toward the Fermi level at the traps. As a result, the electric potential becomes independent of FE polarization, resulting in no potential difference and suppressing the $J_{\mathrm{ON}}/J_{\mathrm{OFF}}$ ratio. This analysis suggests that the balanced point appears just before Fermi-level pinning occurs, where a high $J_{\mathrm{ON}}$ is preserved without degrading the $J_{\mathrm{ON}}/J_{\mathrm{OFF}}$ ratio. As shown in Fig.~3(c), by optimizing the trap condition in the model, despite having identical structural parameters, the TAT-based FTJ shows both improved $J_{\mathrm{ON}}$ and $J_{\mathrm{ON}}/J_{\mathrm{OFF}}$ compared to the conventional FTJ.

\bgroup
\fixFloatSize{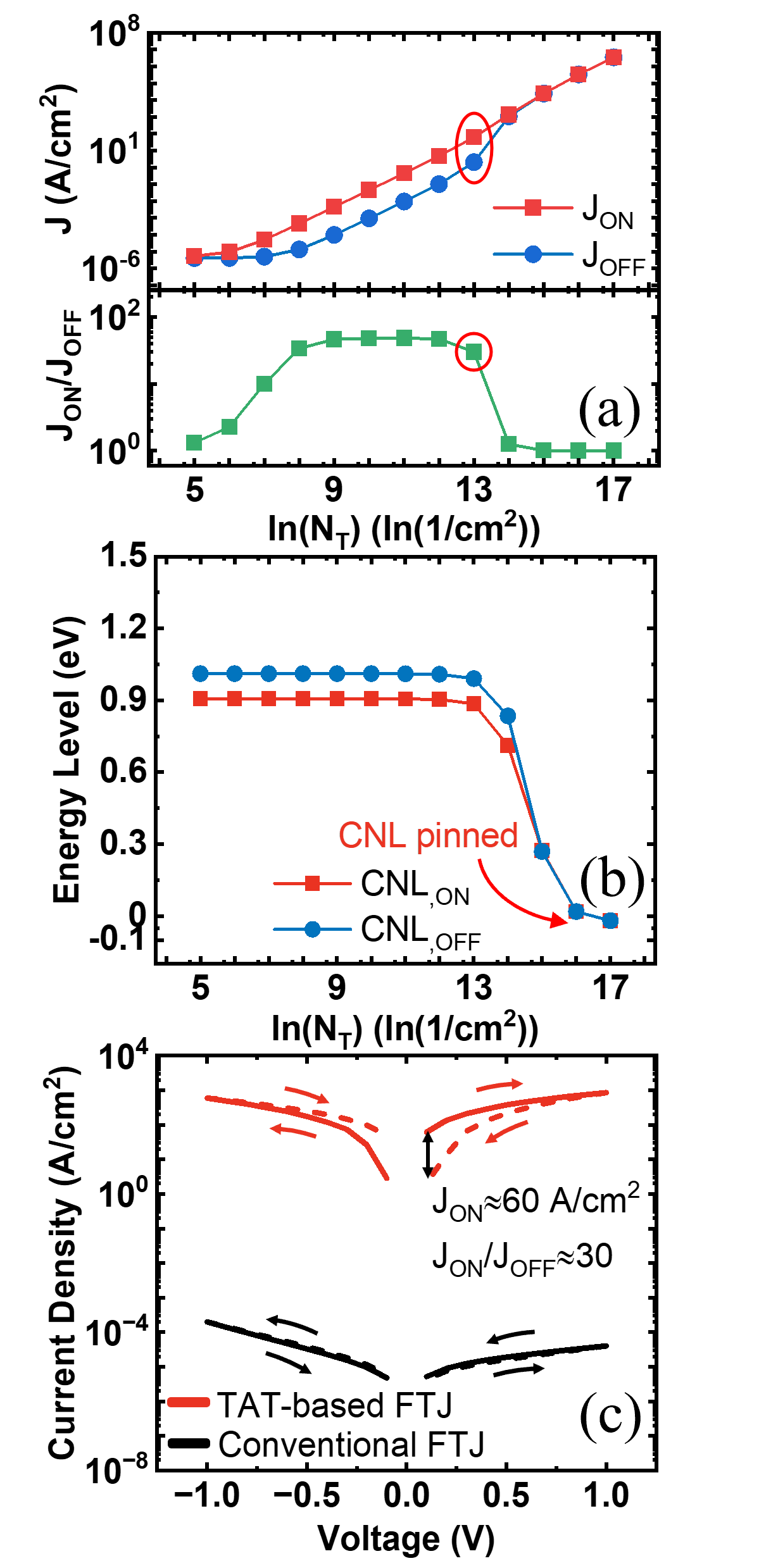}
\begin{figure}[!htbp]
\centering \makeatletter\IfFileExists{images/Figure3.png}{\includegraphics{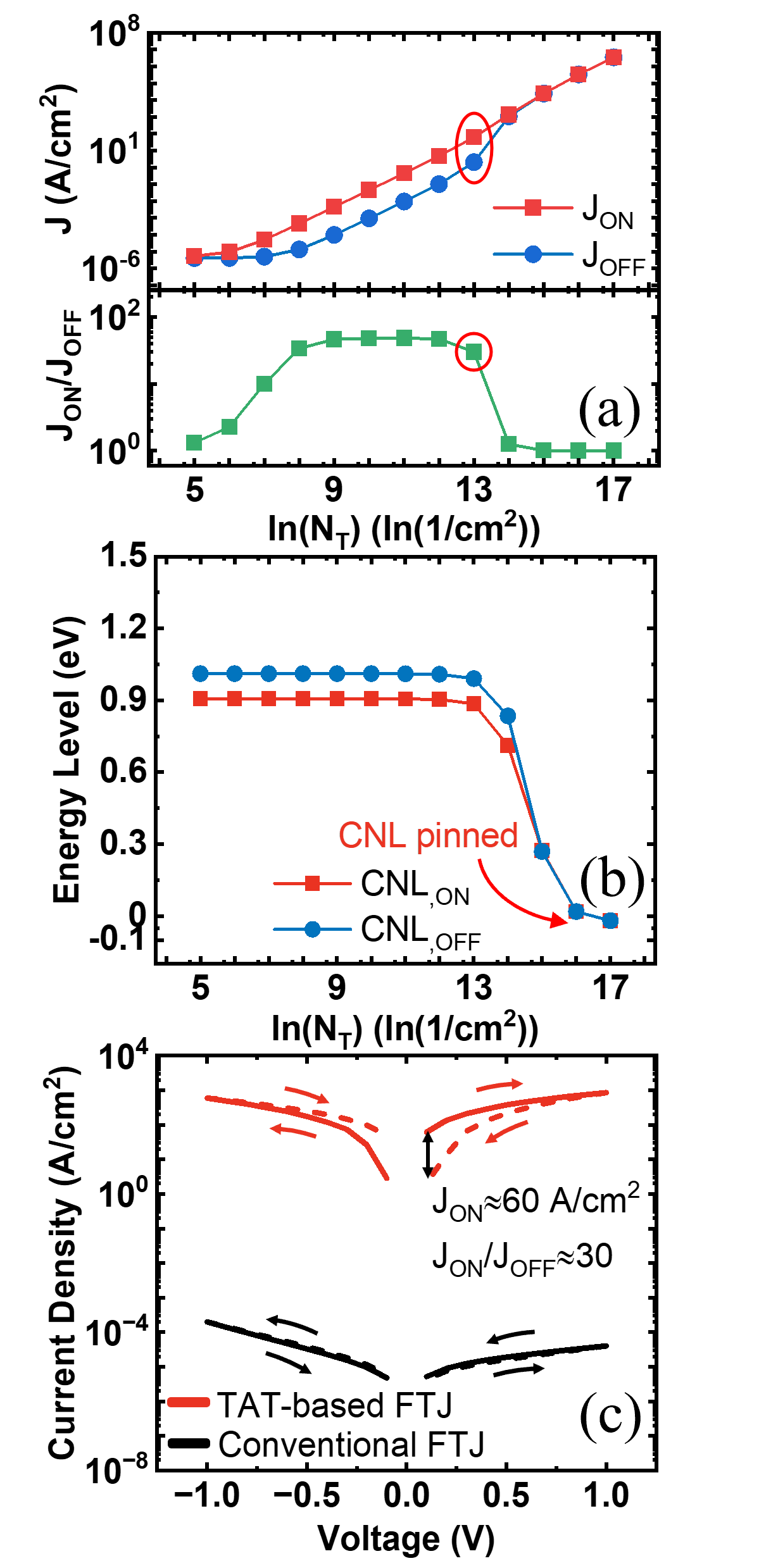}}{\includegraphics{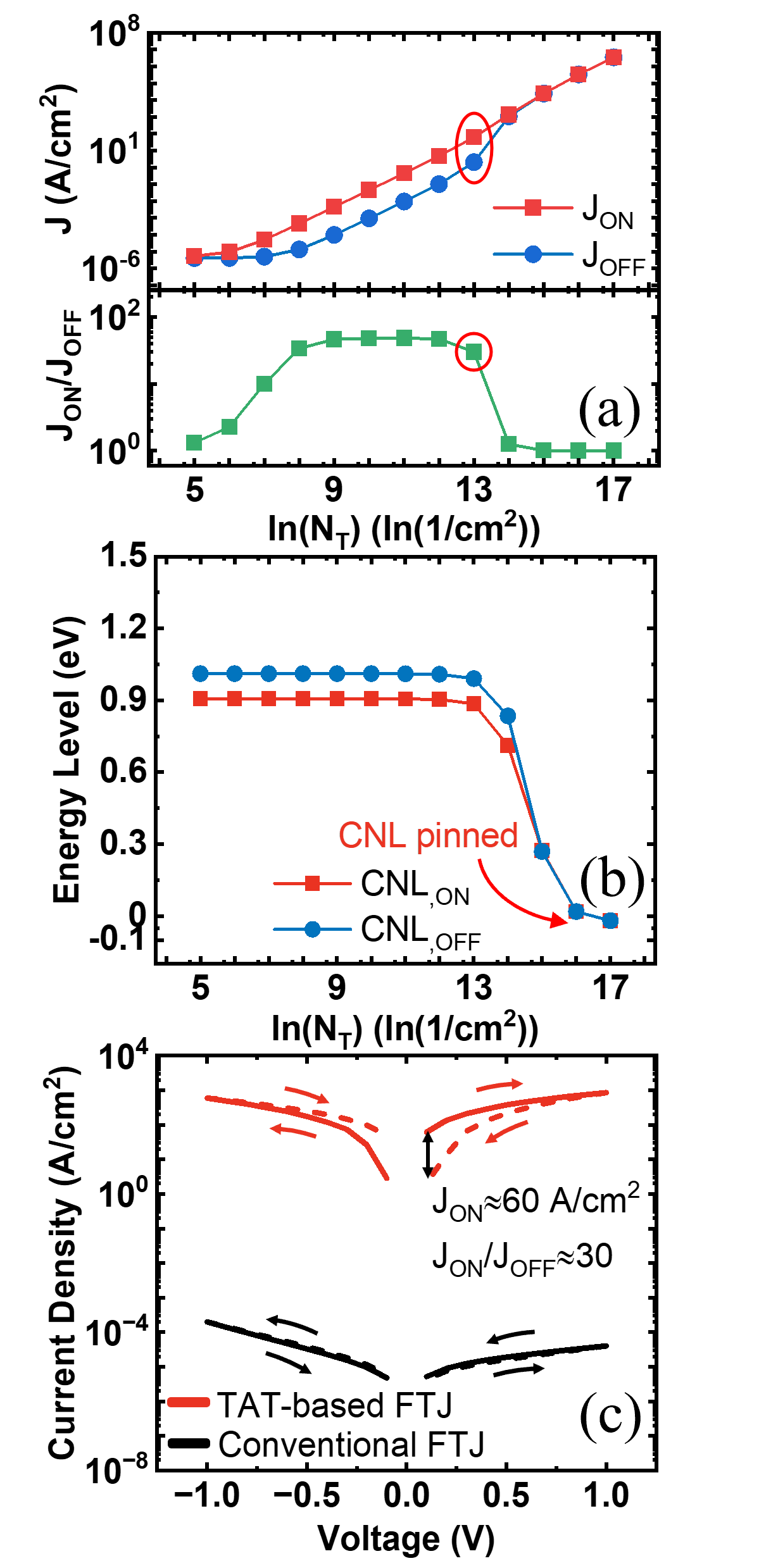}}
\makeatother 
\caption{{\ensuremath{N_{\mathrm{T}}} dependencies of (a) \ensuremath{\mathrm{J}_{\mathrm{TAT}}} and (b) CNL under $V_{\mathrm{write}}=1\;\mathrm{V}$ and $V_{\mathrm{read}}=0.1\;\mathrm{V}$, with \ensuremath{\mathrm{E}_{\mathrm{BE}}} set as the zero-energy reference. (c) J-V curves of the 1.5-nm Al\ensuremath{_{2}}O\ensuremath{_{3}}/3-nm HZO/0.6-nm BIL TAT-based FTJ after trap condition optimization and its conventional counterpart under sweep voltage $\pm1\;\mathrm{V}$.}}
\label{Figure3}
\end{figure}
\egroup

\bgroup
\fixFloatSize{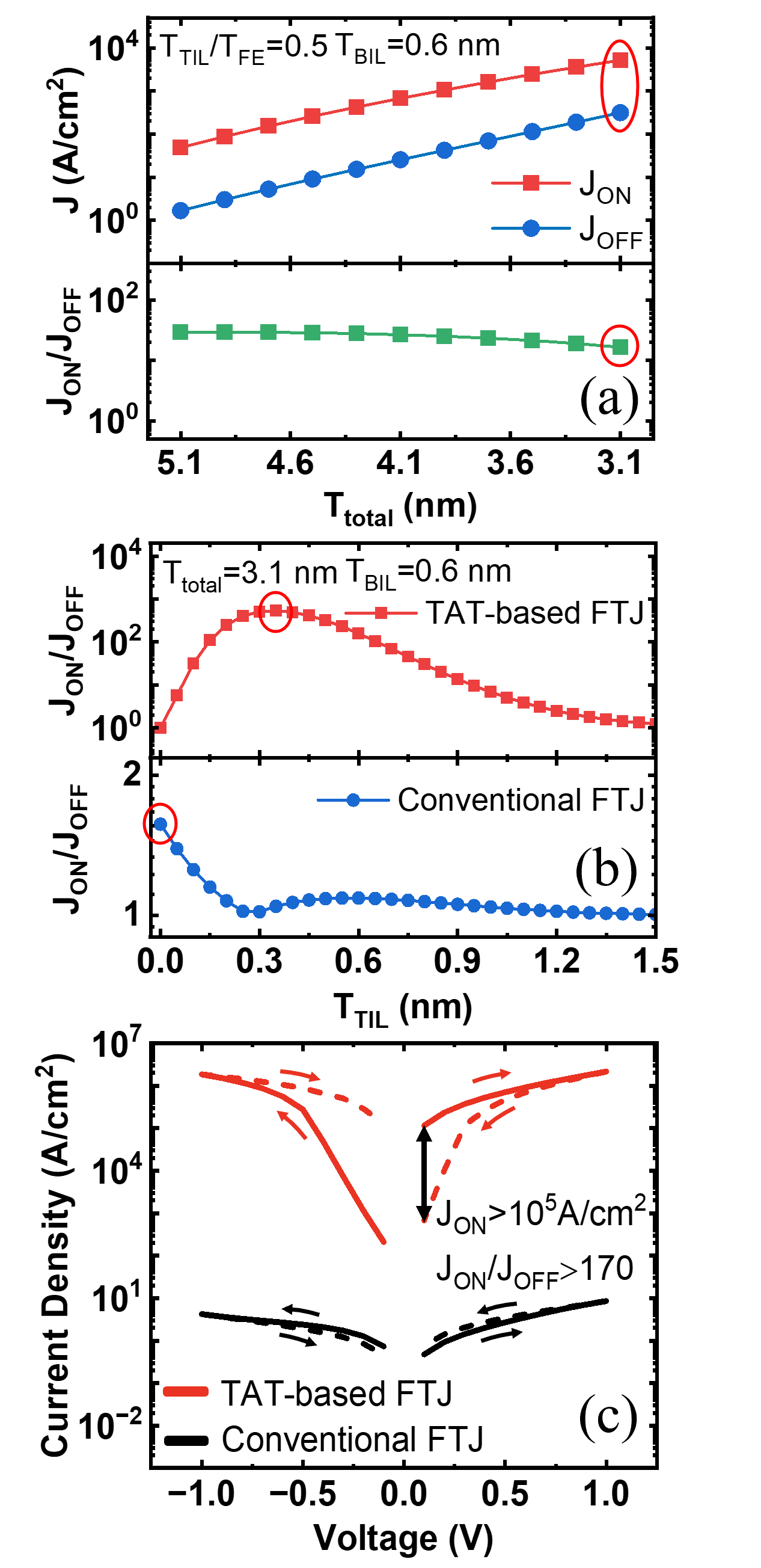}
\begin{figure}[!htbp]
\centering \makeatletter\IfFileExists{images/Figure4.png}{\includegraphics{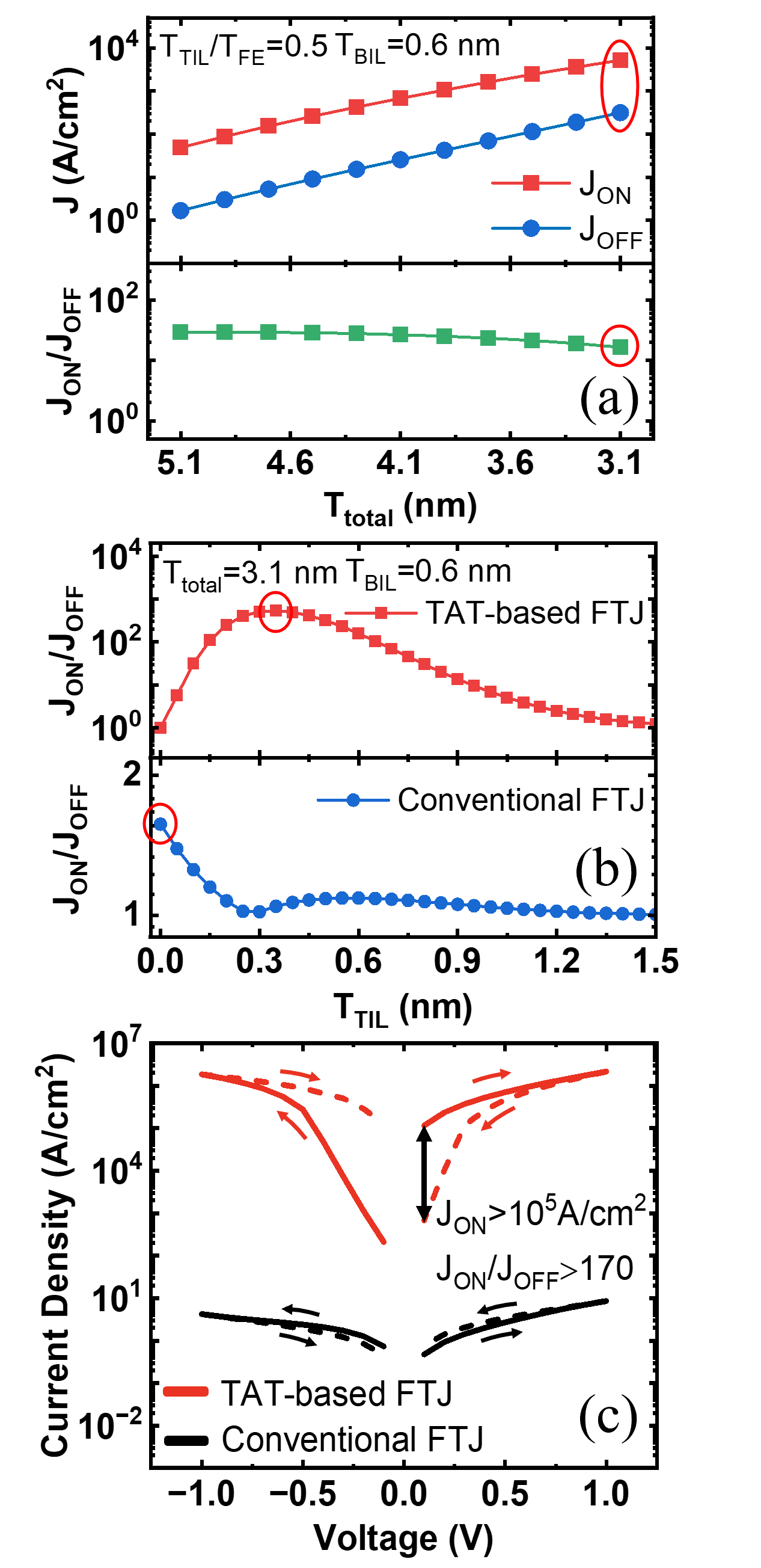}}{\includegraphics{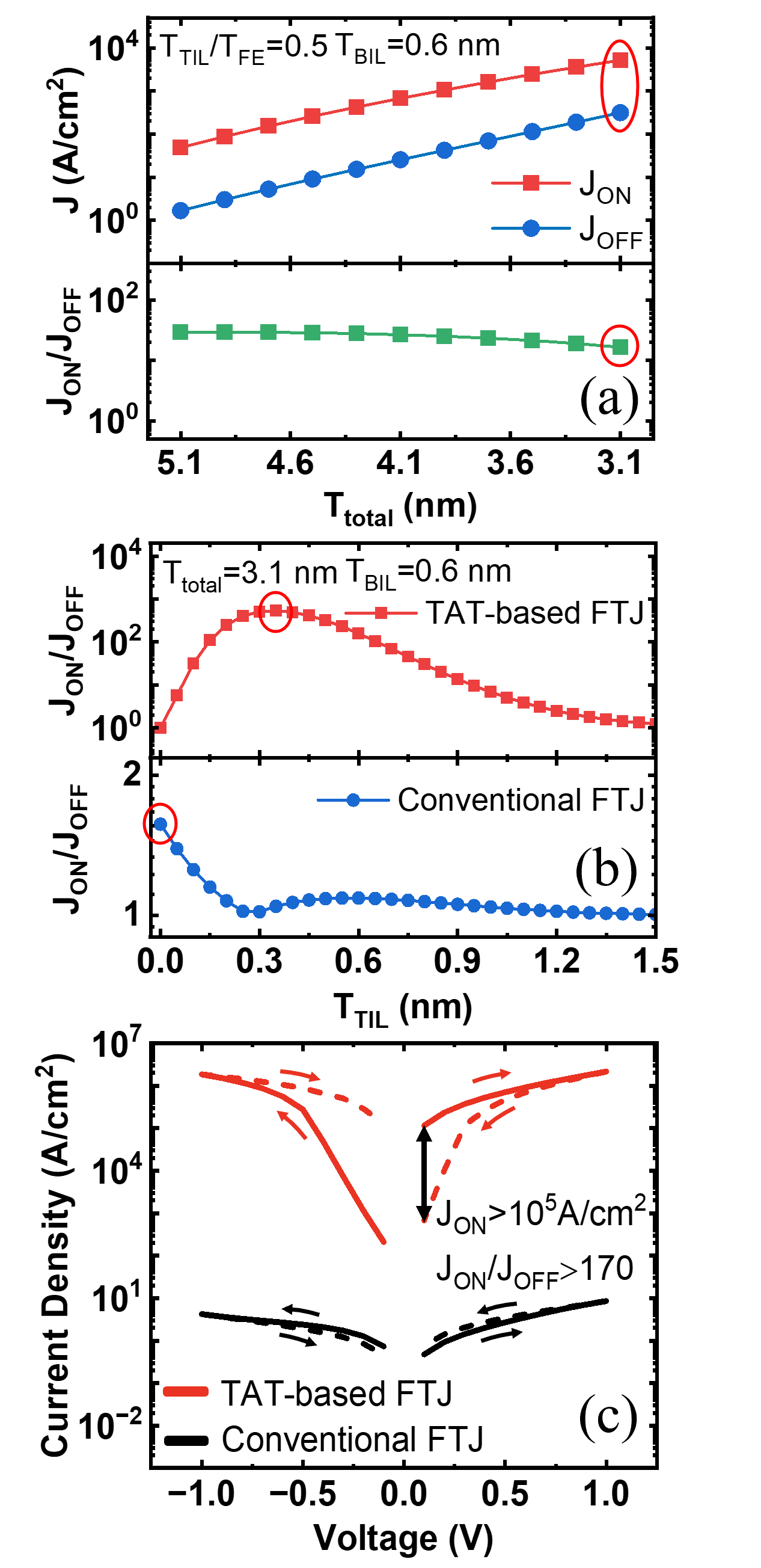}}
\makeatother 
\caption{{(a) Thickness scaling of TAT-based FTJ and (b) thickness proportion modulation of TAT-based FTJ and conventional FTJ under $V_{\mathrm{write}}=1\;\mathrm{V}$ and $V_{\mathrm{read}}=0.1\;\mathrm{V}$. (c) J-V curves of the 0.35-nm Al\ensuremath{_{2}}O\ensuremath{_{3}}/2.15-nm HZO/0.6-nm BIL TAT-based FTJ and the 2.5-nm HZO/0.6-nm BIL conventional FTJ after device structure optimization under sweep voltage $\pm1\;\mathrm{V}$.
}}
\label{Figure4}
\end{figure}
\egroup

Beyond investigating trap conditions, the device structure also plays an important role in meeting the performance requirements of IMC devices. Thickness scaling provides a straightforward strategy to achieve higher $J_{\mathrm{ON}}$. As shown in Fig.~4(a), the total thickness ($T_{\mathrm{total}}$) is scaled down from 5.1~nm to 3.1~nm, maintaining the thickness proportion between the TIL and the FE layer. Under the operation condition of $V_{\mathrm{write}} = 1\ \mathrm{V}$ and $V_{\mathrm{read}} = 0.1\ \mathrm{V}$, the $J_{\mathrm{ON}}$ of the TAT-based FTJ exceeds $10^3$~A/cm$^2$ at $T_{\mathrm{total}} = 3.1$~nm, while the $J_{\mathrm{ON}}$ of the conventional FTJ remains below 1~A/cm$^2$ under the same scaling process (see Sec.~S2 of the supplementary material). To enhance the $J_{\mathrm{ON}}/J_{\mathrm{OFF}}$ ratio, the thickness proportion between the TIL and the FE layer is adjusted by varying the TIL thickness ($T_{\mathrm{TIL}}$) while keeping the total thickness constant. Figure~4(b) illustrates the $J_{\mathrm{ON}}/J_{\mathrm{OFF}}$ ratio of TAT-based and conventional FTJs under different $T_{\mathrm{TIL}}$ values. Notably, the $J_{\mathrm{ON}}/J_{\mathrm{OFF}}$ ratio of the TAT-based FTJ strongly correlates with the potential difference at the TIL–FE interface, as previously discussed. Excessively large $T_{\mathrm{TIL}}$ causes a substantial voltage drop across the TIL and reduces FE polarization due to its dependence on $T_{\mathrm{FE}}$ and the minor loop effect.~\cite{ref24} Conversely, excessively small $T_{\mathrm{TIL}}$ induces a large voltage drop across the FE layer, also reducing the potential difference at the TIL-FE interface. Therefore, the highest $J_{\mathrm{ON}}/J_{\mathrm{OFF}}$ ratio is obtained when $T_{\mathrm{TIL}}$ lies at an intermediate value that balances these competing effects. In contrast, the $J_{\mathrm{ON}}/J_{\mathrm{OFF}}$ ratio of the conventional FTJ primarily depends on the overall energy band variation of the device, where the TIL and BIL exert opposing influences on the switching polarity. Overall, the conventional FTJ shows a limited $J_{\mathrm{ON}}/J_{\mathrm{OFF}}$ ratio even under its most favorable conditions compared to that of TAT-based FTJ.

With the optimized device structure, the trap condition is revisited to achieve the optimal performance. (see Sec.~S3 of the supplementary material). Figure~4(c) compares the TAT-based FTJ and the conventional FTJ, both with optimized device structures. The TAT-based FTJ achieves $J_{\mathrm{ON}} > 10^5\ \mathrm{A/cm^2}$ and $J_{\mathrm{ON}}/J_{\mathrm{OFF}}$ ratio $> 170$ under low voltage operation, both substantially superior to those of the conventional FTJ.

In conclusion, the characteristics of TAT-based FTJ are comprehensively investigated. Theoretical analysis provides guidelines for optimizing trap conditions and device structure to enhance the $J_{\mathrm{ON}}$ and $J_{\mathrm{ON}}/J_{\mathrm{OFF}}$ ratio simultaneously, leveraging the TAT mechanism to overcome the intrinsic bottlenecks of the conventional FTJs. The simulated TAT-based FTJ achieves an ultra-high $J_{\mathrm{ON}} > 10^5\ \mathrm{A/cm^2}$ and a remarkable $J_{\mathrm{ON}}/J_{\mathrm{OFF}}$ ratio exceeding 170. Although the performance attained from actual experiments depends on various material-dependent and fabrication-related factors, the simulation results in this work suggest the strong potential of TAT-based FTJs to outperforming the conventional FTJ and fulfilling the demands of nanoscale IMC. Future efforts incorporating first-principles calculations to obtain more accurate material and trap-related parameters, rather than relying on the values adopted from previous literature in the present simulations, together with examining additional possible conduction pathways, may help provide a more complete understanding of the device mechanism beyond the present analysis.

\section{SUPPLEMENTARY MATERIAL}
The supplementary material provides detailed information on the simulation methods of the FTJ model, analysis of conventional FTJ scaling, and trap condition refinement following structural optimization.

\section{ACKNOWLEDGMENTS}
This study is supported by the National Science and Technology Council of Taiwan under grant: 112-2221-E-A49-164-MY3 and 113-2622-8-A49-012-SB.
    
\section{AUTHOR DECLARATIONS}

\subsection{Conflict of Interest}The authors have no conflicts to disclose.

\subsection{Author Contributions}

\section{DATA AVAILABILITY}
The data that support the findings of this study are available from the corresponding author upon reasonable request.

\bibliography{article}
\end{document}